\documentclass[journal, onecolumn]{IEEEtran}
\usepackage{tikz}
\usetikzlibrary{decorations.pathmorphing, shapes}
\usetikzlibrary{calc}
\usetikzlibrary{arrows}
 \usepackage{relsize}
 \usepackage[americanvoltage]{circuitikz}

\ifCLASSINFOpdf
  
\else

\fi
\usepackage{amsmath}
\usepackage{amssymb}
\usetikzlibrary{calc}
\usepackage[americanvoltage]{circuitikz}
\usepackage{paralist}

\usepackage[caption=false]{subfig}

\newcommand{\vect}[1]{\mathbf{#1}}

\newcommand{\matr}[1]{\mathbf{#1}}

\newcommand{\pref}[1]{(\ref{#1})}
\newcommand{\junk}[1] {}

\def\XXint#1#2#3{{\setbox0=\hbox{$#1{#2#3}{\int}$}
\vcenter{\hbox{$#2#3$}}\kern-.5\wd0}}

\newcommand*\widebar[1]{%
  \hbox{%
    \vbox{%
      \hrule height 0.5pt 
      \kern0.3ex
      \hbox{%
        \kern-0.05em
        \ensuremath{#1}%
        \kern-0.05em
      }%
    }%
  }%
}

\begin{document}

\title{A Novel Single-Source Surface Integral Method to Compute Scattering from Dielectric Objects}
\author{Utkarsh~R.~Patel,~\IEEEmembership{Student~Member,~IEEE}, 
        Piero~Triverio,~\IEEEmembership{Senior Member,~IEEE}, and
Sean V.~Hum,~\IEEEmembership{Senior Member,~IEEE}
\\[12pt]
Submitted to the IEEE Antennas and Wireless Propagation Letters on November 18, 2016
\thanks{Manuscript received ...; revised ...}%
\thanks{U.~R.~Patel, P.~Triverio, and S. V.~Hum are with the Edward S. Rogers Sr. Department of Electrical and Computer Engineering, University of Toronto, Toronto, M5S 3G4 Canada (email: utkarsh.patel@mail.utoronto.ca, piero.triverio@utoronto.ca, sean.hum@utoronto.ca).}
}

\markboth{Submitted to IEEE Antennas and Wireless Propagation Letters on November 18, 2016}{IEEE Antennas and Wireless Propagation Letters}%

\maketitle
\begin{abstract}
Using the traditional surface integral methods, the computation of scattering from a dielectric object requires two equivalent current densities on the boundary of the dielectric.
In this paper, we present an approach that requires only a single current density. 
Our method is based on a surface admittance operator and is applicable to dielectric bodies of arbitrary shape.
The formulation results in four times lower memory consumption and up to eight times lower time to solve the linear system than the traditional PMCHWT formulation.
Numerical results demonstrate that the proposed technique is as accurate as the PMCHWT formulation.
\end{abstract}
\begin{IEEEkeywords}
Equivalence principle, surface admittance operator, integral equation method
\end{IEEEkeywords}

\section{Introduction}

Electromagnetic scattering problems are important for the design of electromagnetic surfaces, radar systems and imaging scanners. 
Since these problems have open boundary conditions, they are ideally analyzed with integral equation methods.
In integral equation methods, a perfect electric conductor (PEC) may be replaced by the surrounding medium and an equivalent electric current density on its surface.
Similarly, a perfect magnetic conductor (PMC) can be modeled with an equivalent magnetic current density on its surface.
Two methods for handling dielectrics exists:  volumetric formulations and surface formulations.
Surface formulations are more efficient than volumetric formulations as they only require surface as opposed to volumetric meshing~\cite{Jin2011}.
However, unlike PECs or PMCs which can be modelled with a single equivalent current density, a dielectric object in most surface formulations is based on a variant of the so-called Poggio-Miller-Chang-Harrington-Wu-Tsai (PMCHWT) formulation~\cite{Poggio1973, Chang1977, Wu1977}. This formulation requires both equivalent electric and magnetic currents because neither the tangential electric field nor the tangential magnetic field is zero on a dielectric's boundary.
With the Schur complement, a single-source formulation can be derived numerically from the PMCHWT formulation~\cite{Qian2007}. However, such a formulation may lead to numerical issues when applied to lossless dielectrics.

In this work, we present a novel approach to compute scattering from dielectric objects using the so-called differential surface admittance operator~\cite{DeZ05}. 
In this formulation, following the equivalence principle, the dielectric is replaced by the surrounding medium and a \emph{contrast} electric current density.
The concept of a differential surface admittance operator was originally proposed in two dimensions for conductors and dielectrics of canonical shapes~\cite{DeZ05}.
Later, the formulation was generalized to 2D objects of arbitrary shape~\cite{TMTT16}.
More recently, a 3D differential surface admittance operator was proposed for cylindrical conductors based on the eigenfunction expansion approach~\cite{Huynen2016}.
In our work, the 3D differential surface admittance operator is derived for dielectrics of arbitrary shape.
For scattering problems, the main advantage of the proposed approach over the PMCHWT formulation is that the proposed approach only requires a single equivalent current density on the boundary, which reduces the computational time and memory consumption required to solve the problem.
In comparison to a numerical derivation based on the Schur complement of PMCHWT formulation~\cite{Qian2007}, the proposed formulation provides more physical insight and may be potentially advantageous when applied to dielectric objects inside a stratified medium since it does not require the computation of reaction integrals due to magnetic currents in the equivalent problem.

\begin{figure}[t]
\centering
\subfloat[\label{fig:original} Original problem] {\begin{tikzpicture}[scale = 1]
\draw [black, fill = black!40] plot [smooth cycle] coordinates {(1.75,0) (1.0, 1.2) (0,1) (-0.5,0.8) (-1,0) (-0.5,-0.9) (0,-1) (0.9, -1.2)};
\draw[black, ->, line width = 0.4mm] (0.5, 1.20) -- (0.63, 0.8 );
\node at (0.63, 0.8) [below] {$\hat{\vect{n}}$};
\node at (0.0,0.5) [left]{${\cal V}$};
\node at (0.8,-0.8) {$({\mu}, {\varepsilon})$};
\node at (0.5,-0.2) {${\cal E}(\vect{r}), {\cal H}(\vect{r})$};
\node at (-0.9,0) [left]{${\cal S}$};
\node at (-0.5,1.8) {$(\vect{E}^{i}, \vect{H}^{i})$};
\draw [->,decorate,decoration=snake] (-0.5,1.6) -- (-0.1,0.97);
\draw [->,decorate,decoration=snake] (1.4,0.8) -- (1.5,1.7);
\node at (1.5,1.8) {$(\vect{E}^{s}, \vect{H}^{s})$};
\node at (-0.9, -1.3) {$\mu_0,\varepsilon_0$};
\end{tikzpicture}}
\subfloat[\label{fig:equivalent} Equivalent problem] {\begin{tikzpicture}[scale = 1]
\draw [black, fill = white] plot [smooth cycle] coordinates {(1.75,0) (1.0, 1.2) (0,1) (-0.5,0.8) (-1,0) (-0.5,-0.9) (0,-1) (0.9, -1.2)};

\draw[black, ->, line width = 0.4mm] (1.8, -0.35) -- (1.30, -1.09 );
\node at (1.6, -0.85) [right] {$\vect{J}_{eq}$};

\node at (0.7,-0.7) {$({\mu}_0, {\varepsilon}_0)$};
\node at (0.5,0.2) {$\widetilde{\cal E}(\vect{r}), \widetilde{\cal H}(\vect{r})$};
\node at (-0.5,1.8) {$(\vect{E}^{i}, \vect{H}^{i})$};
\draw [->,decorate,decoration=snake] (-0.5,1.6) -- (-0.1,0.97);
\draw [->,decorate,decoration=snake] (1.4,0.8) -- (1.5,1.7);
\node at (1.5,1.8) {$(\vect{E}^{s}, \vect{H}^{s})$};
\node at (-0.9, -1.3) {$\mu_0, \varepsilon_0$};
\end{tikzpicture}}
\caption{(a): Original scatterer of arbitrary shape. (b): Equivalent problem with scatterer replaced by the surrounding medium and an equivalent current density.}
\label{fig:configuration}
\end{figure}
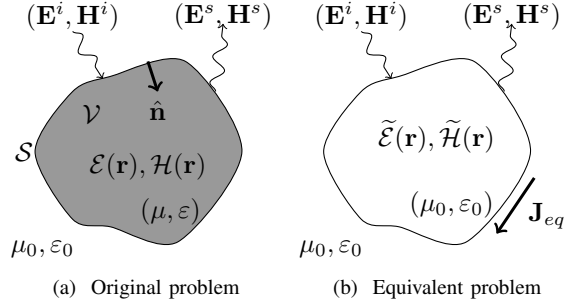

\section{Theoretical Formulation}

\subsection{Original Problem}
\label{sec:original}
We consider scattering from a closed dielectric object of arbitrary shape surrounded by free space, as shown in Fig.~\ref{fig:original}.
The dielectric has permittivity $\varepsilon$, permeability $\mu$, and intrinsic impedance $\eta$.
The volume and the enclosing surface of the scatterer are denoted by~${\cal V}$ and ${\cal S}$, respectively.
The electric and magnetic fields inside ${\cal V}$ in the original problem are given by ${\cal E} (\vect{r})$ and ${\cal H}(\vect{r})$.
Furthermore, we denote the tangential electric and magnetic fields on the boundary in the original problem by $\vect{E}_{t}$ and  $\vect{H}_{t}$, respectively.
The normal cross-product of tangential electric and magnetic fields are expanded as
\begin{align}
\hat{\vect{n}} \times \vect{E}_t &= \sum_{n=1}^{N} e_n \vect{f}_n(\vect{r}) \,, \label{eq:e_expansion} \\
\hat{\vect{n}} \times \vect{H}_t &= \sum_{n=1}^{N} h_n \vect{f}_n(\vect{r}) \,, \label{eq:h_expansion}
\end{align}
where $\vect{f}_n(\vect{r})$ is the $n$-th RWG basis function~\cite{Rao1982}.

On the boundary of the dielectric, the tangential electric and magnetic fields are related by the Stratton-Chu formulation~\cite{Stratton2007} 
\begin{subequations}
\begin{align}
-\frac{1}{2} \hat{\vect{n}} \times \vect{E}_{t} 
-j\omega \mu  \hat{\vect{n}} \times  \left[ {\cal L} \left( \hat{\vect{n}} \times \vect{H}_t\right) \right](\vect{r})& \nonumber \\+ \hat{\vect{n}} \times \left[{\cal K} \left( \hat{\vect{n}} \times \vect{E}_t\right) \right](\vect{r})  =& 0\,, \label{eq:EFIE1}\\
\frac{1}{2} \hat{\vect{n}} \times \hat{\vect{n}} \times \eta \vect{H}_{t} 
-j\omega \varepsilon  \eta \hat{\vect{n}} \times \hat{\vect{n}} \times  \left[ {\cal L} \left( \hat{\vect{n}} \times \vect{E}_t \right) \right](\vect{r})& \nonumber \\
- \hat{\vect{n}} \times \hat{\vect{n}} \times \left[{\cal K} \left(\hat{\vect{n}} \times \eta \vect{H}_t \right) \right](\vect{r})  =& 0\,. \label{eq:MFIE1}
\end{align}
\end{subequations}
Equations~\pref{eq:EFIE1} and~\pref{eq:MFIE1} can also be interpreted as the electric field integral equation (EFIE) and the magnetic field integral equation (MFIE), respectively~\cite{Gibson2014}.
Operators~${\cal K}$ and ${\cal L}$ in~\pref{eq:EFIE1} and~\pref{eq:MFIE1} are given by
\begin{align}
\left[{\cal L} \left( \vect{X} \right) \right](\vect{r}) &= \left[ 1 + \frac{1}{k^2} \nabla \nabla \cdot \right] \iint_{\cal S} G(\vect{r}, \vect{r}') \vect{X}(\vect{r}') dS' \\
\left[ {\cal K} \left( \vect{X} \right) \right](\vect{r}) &= \nabla \times \iint_{\cal S} G(\vect{r}, \vect{r}') \vect{X}(\vect{r}') dS'
\end{align}
where
\begin{equation}
G(\vect{r}, \vect{r}') = \frac{1}{4\pi} \frac{e^{-jk \left|{\vect{r} - \vect{r}'}\right|}}{ \left| \vect{r} - \vect{r}' \right| }
\end{equation}
is the Green's function of the dielectric medium and $k = \omega \sqrt{\mu \epsilon}$ is the wavenumber inside the dielectric. 
Next, we substitute~\pref{eq:e_expansion} and~\pref{eq:h_expansion} into~\pref{eq:EFIE1} and~\pref{eq:MFIE1}, and test the resulting equation with RWG functions to obtain the system of equations
\begin{equation}
\begin{bmatrix} 
\matr{K}_{{e}} & \matr{L}_{{e}} \\
\matr{L}_{{m}} & \matr{K}_{m}
\end{bmatrix}\begin{bmatrix} \vect{E} \\ \vect{H}
\end{bmatrix} = 
\begin{bmatrix}
\vect{0}\\
\vect{0}
\end{bmatrix}
\label{eq:SystemMatrix1}
\end{equation}
where $\vect{E} = \begin{bmatrix} e_1 & \hdots & e_N \end{bmatrix}^T$ and $\vect{H} = \begin{bmatrix} h_1 & \hdots & h_N \end{bmatrix}^T$ contain the expansion coefficients in~\pref{eq:e_expansion} and~\pref{eq:h_expansion}. Matrices $\matr{K}_e$ and $\matr{L}_e$ are generated by testing the  EFIE~\pref{eq:EFIE1} with RWG functions. Similarly, matrices $\matr{K}_m$ and $\matr{L}_m$ are generated by testing the MFIE~\pref{eq:MFIE1} with RWG functions.

From~\pref{eq:SystemMatrix1}, we can obtain an admittance operator $\matr{Y}$ that relates $\vect{H}$ and $\vect{E}$ as
\begin{equation}
\vect{H} = \matr{Y} \vect{E}\,.
\label{eq:Yin}
\end{equation}
While the admittance matrix $\matr{Y}$ can be obtained numerically from the EFIE alone (first equation in~\pref{eq:SystemMatrix1}) or the MFIE alone (second equation in~\pref{eq:SystemMatrix1}), the resulting operator is ill-conditioned due to singularities of the ${\cal K}$ and ${\cal L}$ operator that can lead to interior resonances~\cite{Boeykens2013}. As a remedy to this ill-conditioning, we use a linear combination of the  EFIE and MFIE to obtain the admittance operator~\cite{Gibson2014}. From~\pref{eq:SystemMatrix1}, the surface admittance operator $\matr{Y}$ is
\begin{equation}
\matr{Y} = \left( \alpha \matr{L}_e + (1-\alpha) \matr{K}_m \right)^{-1} \left( \alpha \matr{K}_e + (1-\alpha) \matr{L}_m \right)
\label{eq:DiscreteY}
\end{equation}
where $0 < \alpha < 1$ is the weighting coefficient. Equation~\pref{eq:DiscreteY} requires the solution of a system of size $N$. However, in the presence of many dielectric scatterers, the cost of solving this system is very small compared to the gains of solving a system with half as many total number of unknowns as the PMCHWT formulation.

\subsection{Equivalent Problem}

Next, we replace the scatterer by the material of the surrounding medium, as shown in Fig.~\ref{fig:equivalent}. 
The electric and magnetic fields inside ${\cal S}$ in this equivalent configuration are given by $\widetilde{\cal E} (\vect{r})$ and $\widetilde{\cal H} (\vect{r})$. 
Furthermore, we impose that the electric field on the boundary ${\cal S}$ in the equivalent problem is the same as the electric field on ${\cal S}$ in the original problem in Fig.~\ref{fig:original}. Hence, the tangential electric and magnetic fields on the boundary in the equivalent configuration are given by $\vect{E}_{t}$ and $\widetilde{\vect{H}}_{t}$\footnote{It is important to note that this formulation is different than the PMCHWT formulation which enforces null fields inside the dielectric for the exterior problem and null fields in the surrounding medium for the interior problem.}. 
The normal cross-product of the tangential magnetic field in the equivalent configuration is also discretized with RWG functions,
\begin{equation}
\hat{\vect{n}} \times \widetilde{\vect{H}}_t = \sum_{n=1}^{N} \tilde{h}_n \vect{f}_n(\vect{r})\,,
\label{eq:h2_expansion}
\end{equation}
and their expansion coefficients are stored in vector $\widetilde{\vect{H}} = \begin{bmatrix} \tilde{h}_1 & \hdots & \tilde{h}_N \end{bmatrix}^T$.

In the equivalent problem, $\vect{E}_t$ and $\widetilde{\vect{H}}_t$ satisfy the EFIE~\pref{eq:EFIE1} and MFIE~\pref{eq:MFIE1} but with the material parameters and Green's function of the surrounding medium. 
Therefore, we can obtain the admittance operator $\widetilde{\matr{Y}}$ for the equivalent problem such that
\begin{equation}
\widetilde{\vect{H}} = \widetilde{\matr{Y}} \vect{E}\,.
\label{eq:Yout}
\end{equation}
Similar to Sec.~\ref{sec:original}, we numerically obtain $\widetilde{\matr{Y}}$ by first substituting~\pref{eq:e_expansion} and~\pref{eq:h2_expansion} into~\pref{eq:EFIE1} and~\pref{eq:MFIE1}, and then testing the resulting equation with RWG functions. 
Based on the result of~\pref{eq:DiscreteY}, the discretized admittance operator $\widetilde{\vect{Y}}$ for the equivalent configuration is 
\begin{equation}
\widetilde{\matr{Y}} = \left( \alpha \widetilde{\matr{L}}_{{e}} + (1-\alpha) \widetilde{\matr{K}}_{{m}} \right)^{-1} \left( \alpha \widetilde{\matr{K}}_{{e}} + (1-\alpha) \widetilde{\matr{L}}_{{m}} \right)\,,
\label{eq:DiscreteY2}
\end{equation}
where $\widetilde{\matr{L}}_{{e}}$, $\widetilde{\matr{L}}_{{m}}$, $\widetilde{\matr{K}}_{{e}}$, and $\widetilde{\matr{K}}_{{m}}$ are the same as  matrices defined in~\pref{eq:SystemMatrix1}, but with the material properties of the surrounding medium.

In order to restore the fields outside ${\cal S}$ in the equivalent problem, we introduce an equivalent \emph{contrast} current density 
\begin{equation}
\vect{J}_{eq}(\vect{r}) = \sum_{n=1}^{N} j_n \vect{f}_n(\vect{r})\,
\label{eq:Jeq1}
\end{equation}
on ${\cal S}$. The expansion coefficients in~\pref{eq:Jeq1} are collected into vector $\vect{J} = \begin{bmatrix}  j_1 & \hdots & j_N \end{bmatrix}^T$.
From the equivalence principle~\cite{Gibson2014}, this current density is given by
\begin{equation}
\vect{J}_{eq}(\vect{r}) = \hat{\vect{n}} \times \left[\widetilde{\vect{H}}_{t}(\vect{r}) - \vect{H}_{t}(\vect{r}) \right]\,.
\label{eq:Jeq}
\end{equation}
By substituting \pref{eq:h_expansion},  \pref{eq:h2_expansion}, and \pref{eq:Jeq1} into \pref{eq:Jeq} we obtain the following numerical relationship
\begin{equation}
\vect{J} = \left( \widetilde{\vect{H}} - \vect{H} \right)\,.
\label{eq:Jeq2}
\end{equation}
Since the electric field on the boundary ${\cal S} $ is the same in the original and the equivalent configuration, we do not need to introduce a magnetic equivalent current density on ${\cal S}$. 
The absence of the magnetic equivalent current density leads to half as many unknowns as the PMCHWT method, which ultimately results in lower memory requirements and faster computations.

\subsection{Differential Surface Admittance Operator}

Next we substitute~\pref{eq:Yin} and~\pref{eq:Yout} into~\pref{eq:Jeq2} to obtain
\begin{equation}
\vect{J} = \underbrace{\left[\widetilde{\matr{Y}} - \matr{Y}  \right]}_{\matr{Y}_s} \vect{E} \,,
\label{eq:Ys1}
\end{equation} 
where the term inside the square bracket can be interpreted as a differential surface admittance operator $\matr{Y}_s$. 
The proposed technique to derive~\pref{eq:Ys1} is valid for objects of arbitrary shape, as opposed to the eigenfunction method, which gives~\pref{eq:Ys1} only for canonical objects such as cylinders~\cite{Huynen2016} for which the eigenfunctions of the wave equation are known analytically.

\subsection{Multiple Scatterers}

In the presence of multiple scatterers, we replace all scatterers by the surrounding medium and introduce an equivalent electric current density on the surface of each scatterer. 
Furthermore, the surface operator for each scatterer is obtained. Since the surface operator of each scatterer is independent of other scatterers, this step can be parallelized efficiently.

\subsection{Solution of the Scattering Problem}

Following the equivalence, we now have a homogeneous medium with the material parameters of free space, and an equivalent current density on the surface of each scatterer. 
Next, we relate the equivalent current density to the incident electric and magnetic fields through the EFIE and the MFIE.
For the equivalent problem, the equivalent current density relates the tangential electric field on ${\cal S}$ by
\begin{equation}
\hat{\vect{n}} \times \vect{E}_t(\vect{r}) = -j\omega \mu \hat{\vect{n}} \times  \left[{\cal L} \vect{J}_{eq} \right](\vect{r}) + \hat{\vect{n}} \times  \vect{E}^i(\vect{r})\,,
\label{eq:EFIE2}
\end{equation}
and the tangential magnetic field on ${\cal S}$ by
\begin{equation}
\hat{\vect{n}} \times \hat{\vect{n}} \times \widetilde{\vect{H}}_t(\vect{r}) = \hat{\vect{n}} \times \hat{\vect{n}} \times \left[ {\cal K} \vect{J}_{eq}  \right] (\vect{r}) + \hat{\vect{n}} \times \hat{\vect{n}} \times \vect{H}^{i}(\vect{r})\,,
\label{eq:MFIE2}
\end{equation}
where $\vect{E}^{i}(\vect{r})$ and $\vect{H}^{i}(\vect{r})$ represent the incident electric and magnetic fields, respectively. 
Next, we substitute~\pref{eq:Jeq1},~\pref{eq:h2_expansion} and~\pref{eq:e_expansion} into~\pref{eq:EFIE2} and~\pref{eq:MFIE2}, and then test the resulting equation with RWG functions to obtain
\begin{subequations}
\begin{align}
\matr{D} \vect{E} &= \matr{L}_{out} \vect{J} + \vect{V}_e\,, \label{eq:EFIE3}\\
\widetilde{\matr{D}} \widetilde{\vect{H}} &= \matr{K}_{out} \vect{J} + \vect{V}_m\,, \label{eq:MFIE3}
\end{align}
\end{subequations}
where matrices $\matr{D}$, $\widetilde{\matr{D}}$, $\matr{L}_{out}$, and $\matr{K}_{out}$ are generated by testing~\pref{eq:EFIE2} and~\pref{eq:MFIE2} by RWG functions. The vectors $\vect{V}_e$ and $\vect{V}_h$ are generated by testing the  incident tangential electric and magnetic fields by the RWG functions.

Next, we substitute \pref{eq:Ys1} and~\pref{eq:Yout} into~\pref{eq:EFIE3} and~\pref{eq:MFIE3} to eliminate~$\widetilde{\vect{H}}$ and $\vect{J}$ and obtain
\begin{subequations}
\begin{align}
\matr{D} \vect{E} &= \matr{L}_{out} \matr{Y}_s \vect{E} + \vect{V}_e \label{eq:EFIE4} \\
\widetilde{\matr{D}}\widetilde{\vect{Y}} \vect{E} &= \matr{K}_{out} \matr{Y}_s \vect{E} + \vect{V}_h\,. \label{eq:MFIE4}
\end{align}
\end{subequations}
We can now solve for~$\vect{E}$ and then obtain the equivalent current $\vect{J}$ through~\pref{eq:Ys1}.
In order to avoid numerical issues such as interior resonances~\cite{Boeykens2013}, we take average of both~\pref{eq:EFIE4} and~\pref{eq:MFIE4}, and then solve for $\vect{E}$. 
Once $\vect{E}$ and $\vect{J}$ are found, the scattered field can be easily calculated.
Note that the size of all matrices in~\pref{eq:EFIE4} and~\pref{eq:MFIE4} is $N \times N$. The PMCHWT formulation for this problem would require assembling and solving matrices of size $2N \times 2N$. 
Therefore, the proposed formulation leads to four times lower memory requirements and approximately eight times the computational speed-up in solving the system than the PMCHWT formulation.

\section{Numerical Results}
\label{sec:Results}

\subsection{Dielectric Sphere}
\label{sec:OneSphere}

\begin{figure}[t]
\centering
\includegraphics[width=0.5\columnwidth]{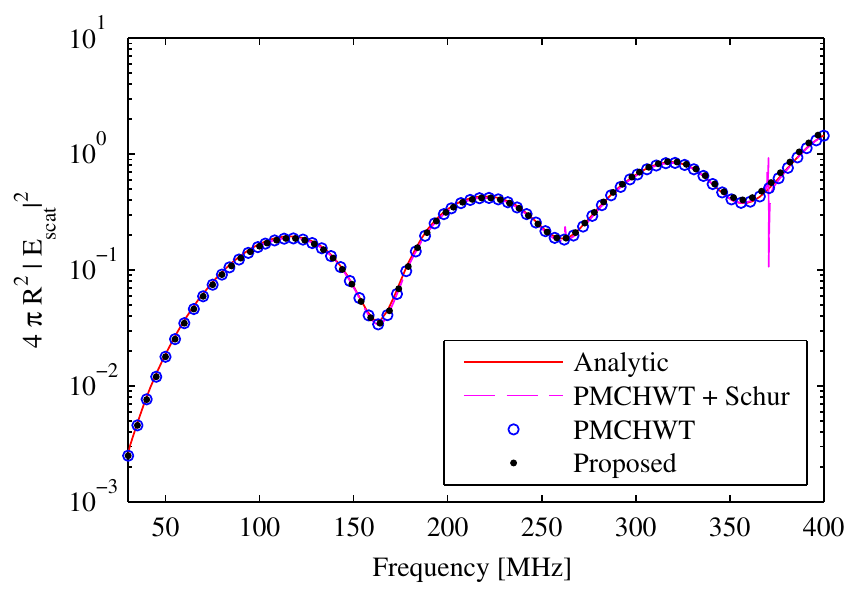}
\caption{Scattered electric field in the example considered in Sec.~\ref{sec:OneSphere} calculated with analytic formulas, the PMCHWT formulation, the PMCHWT formulation with Schur complement reduction, and the proposed method.}
\label{fig:RCS_OneSphere}
\end{figure}

To demonstrate the accuracy of the proposed approach, we consider the scenario of a plane wave incident upon a dielectric sphere with relative permittivity $\varepsilon_r =  2.25$ and radius $R=0.5~{\rm m}$.
The sphere is discretized with $1220$ triangular elements. 
In the proposed approach, a total of $1830$ RWG basis functions~\cite{Rao1982} were used to expand the equivalent electric current density.
The PMCHWT formulation required discretization of both electric and magnetic current density, each with $1830$ RWG basis functions.
We calculated the scattered field with the analytic method based on the solution of fields in spherical coordinates~\cite{Har61}, the PMCHWT formulation~\cite{Gibson2014}, the PMCHWT formulation with Schur complement~\cite{Qian2007}, and the proposed method over the frequency range 30-400~MHz.
Figure~\ref{fig:RCS_OneSphere} shows an excellent agreement between all four methods, which validates the accuracy of the proposed method over a wide frequency range. 
While the Schur complement approach works well for conductors~\cite{Qian2007}, it suffers from numerical resonances when applied to lossless dielectrics, as seen in Fig.~\ref{fig:RCS_OneSphere} at 370~MHz and 260~MHz.
The proposed method, on the other hand, does not suffer from such numerical resonances and is more robust.
In addition, if the surrounding medium were a stratified medium, the proposed approach would be computationally more efficient than the Schur complement based reduction technique because it does not require evaluation of reaction integrals due to the magnetic currents, which are expensive to calculate.

\subsection{4 $\times$ 4 Array of Dielectric Spheres}
\label{sec:4by4array}

\begin{figure}[t]
\centering
\includegraphics[width=0.5\columnwidth]{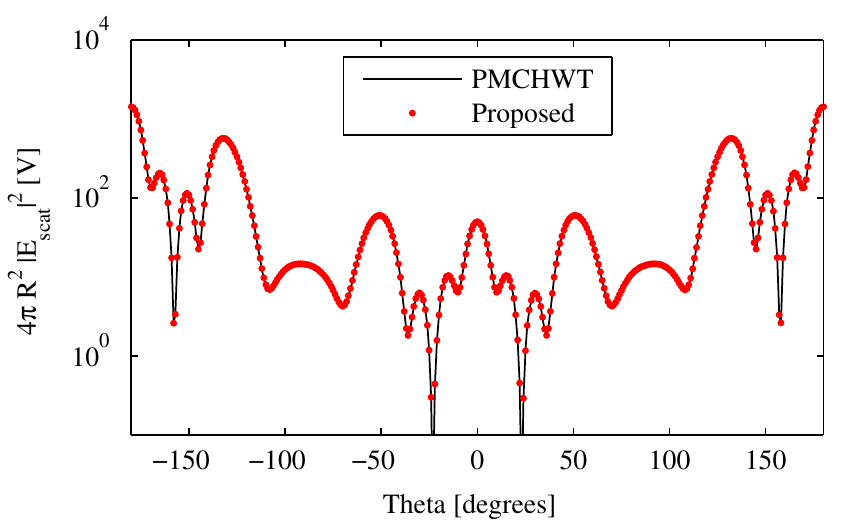}\\
\includegraphics[width=0.5\columnwidth]{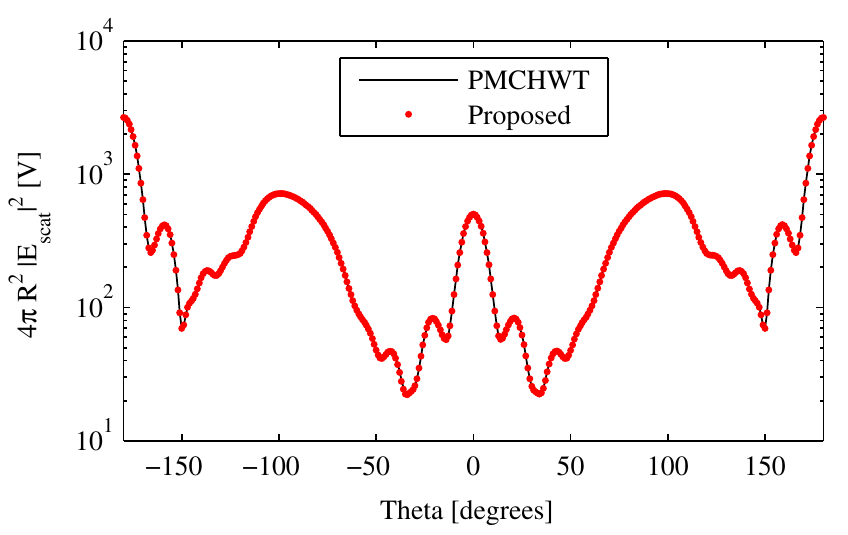}
\caption{Scattered electric field in the example considered in Sec.~\ref{sec:4by4array} obtained with PMCHWT formulation and the proposed method for $\phi = 0^\circ$ (top panel) and $\phi = 45^\circ$ (bottom panel) cuts.}
\label{fig:RCS_16Spheres}
\end{figure}

Next, we consider a $4 \times 4$ array of dielectric spheres placed in the $x$-$y$ plane. 
The array is uniformly spaced along the $x$- and $y$-direction with inter-element spacing of $d_x = d_y = 2~{\rm m}$. 
Geometrical and material properties of each sphere are the same as in Sec.~\ref{sec:OneSphere}.
The array of dielectric spheres was discretized with $9,872$ triangular elements. 
The electric and magnetic equivalent current densities in the PMCHWT formulation and the equivalent electric current density in the proposed method were each discretized with $14,808$ RWG elements.
Figure~\ref{fig:RCS_16Spheres} shows the scattered electric field as a function of elevation angle for $\phi = 0$ and $\phi = \pi/4$ when the array is excited by a $-z$-directed plane wave at $f = 200~{\rm MHz}$. 
It is evident that the proposed method is accurate when compared against the PMCHWT method. 
The computational times and memory requirements for both techniques are summarized in Table~\ref{tab:cputimes}.
The proposed method and the PMCHWT formulation both required almost the same amount of time to generate the system matrix. 
The matrix fill time includes the time required to generate the surface admittance operator.
The solution time for the proposed method, however, is almost 8 times faster than the PMCHWT approach because the proposed approach does not require a magnetic equivalent current density.
The proposed approach also requires four times less memory than the PMCHWT approach.

\begin{table}[t]
\centering
\caption{Computational Time and Memory Requirements for the Example in Sec.~\ref{sec:4by4array}}
\begin{tabular}{|c|c|c|}
\hline 
& {\bf PMCHWT} & {\bf Proposed} \\ \hline
Matrix-fill time (s) & $467$ & $531$ \\ \hline
Solution time (s)  &  $1278$ & $168$ \\ \hline
Total time (s)  &  $1745$ & $699$ \\ \hline
Memory used & $13.08$~GB & $3.47$~GB\\ \hline
\end{tabular}
\label{tab:cputimes}
\newline
\newline
{All computations were performed on a system\\ with a 2.5 GHz CPU, 8 cores, and 16~GB of memory.}
\end{table}

\section{Conclusions}
We have presented a technique to compute scattering from dielectric objects of arbitrary shape using a single equivalent current density. 
The technique replaces dielectrics by the surrounding medium and equivalent electric current densities on the boundary of the dielectrics.
Furthermore, the equivalent current density is related to the electric field on the boundary through the surface admittance operator that is generated numerically.
The proposed method requires half the number of unknowns required by the PMCHWT method which leads to lower computation time and memory consumption.

\bibliographystyle{IEEEtran}
\bibliography{IEEEabrv,biblio3D}

\end{document}